\documentclass[11pt]{article}

\usepackage{graphicx}
\usepackage{natbib}

\usepackage{amssymb}
\usepackage{amsmath,amsfonts,amsbsy}
\usepackage{multirow}

\newsavebox{\astrutbox}
\sbox{\astrutbox}{\rule[-5pt]{0pt}{20pt}}

\def\pr{{\partial}}
\def\eps{{\epsilon}}
\def\bx{{\bf x}}

\def\bv{{\bf v}}

\def\bV{{\bf V}}
\def\be{{\bf e}}

\def\bF{{\bf F}}

\begin{document}

\centerline{\LARGE\bf Steady streaming in a channel with permeable walls}

\centerline{}

\centerline{\large\bf Konstantin Ilin\footnote{Email address: konstantin.ilin@york.ac.uk}}

\centerline{}

\centerline{Department of Mathematics, University of York,
Heslington, York, YO10 5DD, UK}

\begin{abstract}
We study steady streaming in a channel between two parallel permeable walls induced by oscillating (in time) blowing/suction
at the walls. We obtain an asymptotic expansion of the solution of the Navier-Stokes equations in the limit when the amplitude of the normal
displacements of fluid particles near the walls is much smaller that both  the width of the channel and the thickness of the Stokes
layer. It is demonstrated that the magnitude of the steady streaming is much bigger than the corresponding quantity in the case of the steady streaming produced by vibrations of impermeable boundaries.
\end{abstract}


\setcounter{equation}{0}
\renewcommand{\theequation}{1.\arabic{equation}}

\section{Introduction}
\label{Sec1}
It is well-known that an oscillating (in time) body force or vibrations of
the boundary of a domain occupied by a viscous
fluid can produce not only an oscillating flow but also a (relatively) weak steady flow, which is usually
called steady streaming \citep[see][]{Lighthill, Riley1967, Riley2001}. In this paper, we present a theory
of steady streaming in a channel with fixed but permeable walls produced by given velocity
at the walls which is oscillating
in time with angular frequency $\omega$.
The basic parameters of the problem
are the inverse Strouhal number $\epsilon^2$ and the
inverse Reynolds number $\nu$, defined by
\begin{equation}
\epsilon^2=\frac{1}{St}=\frac{V_{0}^{*}}{\omega d}, \quad
\nu=\frac{1}{Re}=\frac{\nu^{*}}{V_{0}^{*}d}
 \label{1.1}
\end{equation}
where $V_{0}^{*}$ is the amplitude of the oscillating velocity at the
walls, $d$ is the distance between the walls and $\nu^{*}$ is the kinematic
viscosity of the fluid.
Parameter $\epsilon^2$ measures the ratio of the amplitude of the
displacements of fluid particles in an oscillating velocity field with amplitude $V_{0}^{*}$ to the distance between the walls.
Another dimensionless parameter which is widely used in literature is the `streaming Reynolds
number' $R_{s} = V_{0}^{*2}/\omega \nu^{*}$. In terms of
parameters $\eps$ and $\nu$,  $R_{s}=\eps^2/ \nu$.
We are interested in the asymptotic behaviour of solutions of the Navier-Stokes equations
in the limit $\eps\ll 1$ and $\nu\sim 1$. This means that
the amplitude of displacements of fluid particles is much smaller than
the thickness of the Stokes layer.
Note that this limit corresponds to small $R_{s}$ ($R_{s}\sim \eps^2 \ll 1$).

Early studies of the steady streaming in a channel induced by vibrations of the walls had been focused
on the problem of peristaltic
pumping in channels and pipes under the assumption of low Reynolds numbers ($R\ll 1$)
and small amplitude-to-wavelength ratio \citep[see, e.g.,][]{Jaffrin, Wilson}. In recent years, there had been considerable renewed interest in the problem motivated by possible applications of steady streaming to
micro-mixing \citep{Selverov, Yi, Carlsson} and to drag reduction in channel flows \citep{Hoepffner}.
In all asymptotic theories of the steady streaming produced by vibrating impermeable boundaries, the magnitude of
steady streaming is $O(\eps^2)$ for $\eps\ll 1$. This is because steady streaming results from the nonlinearity
in the Navier-Stokes equations and parameter $\eps^2$ measures the magnitude of the nonlinear term.
The aim of the present study is to show that if the boundary is permeable, then the magnitude of
steady streaming is $O(\eps)$ for small $\eps$, i.e. much bigger than in the case of an impermeable boundary.

To obtain
an asymptotic expansion,
we employ the Vishik-Lyusternik
method \citep[see, e.g.,][]{Trenogin, Nayfeh}.
It is described in detail in \citet{IS2010}.

The outline of the paper is as follows. In Section 2, we formulate the mathematical
problem. In Section 3, the asymptotic expansion of the solution is described.
In Section 4, we consider simple examples in which the leading order asymptotic solution
can be obtained analytically.
Finally, conclusions are presented in Section 5.


\setcounter{equation}{0}
\renewcommand{\theequation}{2.\arabic{equation}}

\section{Formulation of the problem}

We consider a two-dimensional viscous incompressible flow in an infinite channel
of width $d$. The walls of the channel are permeable for the fluid, and the flow is
produced by a given velocity at the walls which is assumed to be periodic along the channel
with period $L^*$ and oscillating in time with angular frequency $\omega$.
We will use
the following non-dimensional quantities:
\[
\tau=\omega t^{*}, \quad \bx=\frac{\bx^{*}}{d}, \quad
\bv=\frac{\bv^{*}}{V_{0}^{*}}, \quad
p=\frac{p^{*}}{\rho d \omega V_{0}^{*}}.
\]
Here $t^*$ is time; $\bx^*=(x^*,y^*)$; $x^*$ and $y^*$ are Cartesian coordinates, the $x^*$ axis being parallel to the channel;
$\bv^*=(u^*,v^*)$ is the velocity of the fluid; $p^*$ is the pressure; $\rho$ is the density;
$V_{0}^{*}$ is the maximum of the given velocity at the walls over all $x^*$ and $t^*$.
In these variables, the Navier-Stokes equations take the form
\begin{equation}
\bv_{\tau}+\eps^2 (\bv\cdot\nabla)\bv= -\nabla p +\eps^2 \nu \nabla^{2}\bv, \quad \nabla\cdot\bv=0, \label{1}
\end{equation}
where the dimensionless parameters $\eps^2$ and $\nu$ are defined by Eq. (\ref{1.1}).
Equations (\ref{1}) are to be solved subject to the boundary conditions
\begin{equation}
\bv\vert_{y=0} = \bV^{a}(x ,\tau,\eps), \quad
\bv\vert_{y=1} = \bV^{b}(x,\tau,\eps). \label{4}
\end{equation}
Here $\bV^{a}=(U^a,V^a)$
and $\bV^{b}=(U^b,V^b)$ are given functions which are $2\pi$-periodic in $\tau$ and
have zero mean value:
\begin{equation}
\bar{\bV}^{a}\equiv\frac{1}{2\pi}\int\limits_{0}^{2\pi}\bV^{a}(x,\tau)\, d\tau=0, \quad
\bar{\bV}^{b}\equiv\frac{1}{2\pi}\int\limits_{0}^{2\pi}\bV^{b}(x,\tau)\, d\tau=0. \label{2}
\end{equation}
They are also periodic in $x$ with period $L=L^*/d$ and satisfy the condition
\begin{equation}
\int\limits_{0}^{L}V^{a}(x,\tau)\, dx=\int\limits_{0}^{L}V^{b}(x,\tau)\, dx
\label{3}
\end{equation}
which follows from incompressibility of the fluid.
In what follows we are interested in
the asymptotic behaviour of periodic (both in $\tau$ and $x$) solutions of Eqs. (\ref{1}), (\ref{4})
in the limit $\eps\to 0$ and $\nu=O(1)$.
We assume that $\bV^{a,b}(x,\tau,\eps)$ can be presented in the form
\begin{equation}
\bV^{a,b}(x,\tau,\eps)=\bV^{a,b}_{0}(x,\tau)+\eps \, \bV^{a,b}_{1}(x,\tau)
+\eps^2 \bV^{a,b}_{2}(x,\tau)+\dots    \label{5}
\end{equation}
We seek a solution  of (\ref{1}), (\ref{4}) in the form
\begin{equation}
u=u^{r}+u^{a}+u^{b},  \quad
v=v^{r}+\eps \, v^{a}+\eps \, v^{b}, \quad
p=p^{r}+ p^{a}+ p^{b}. \label{6}
\end{equation}
Here $u^{r}$, $v^{r}$, $p^{r}$ are functions of $x$, $y$, $\tau$ and $\eps$;
$u^{a}$, $v^{a}$, $p^{a}$ depend on $x$, $\tau$, $\eps$ and the boundary layer variable
$\xi=y/\epsilon$;
$u^{b}$, $v^{b}$, $p^{b}$ depend on $x$, $\tau$, $\eps$ and the boundary layer variable $\eta=(1-y)/\epsilon$.
Functions $u^{r}$, $v^{r}$, $p^{r}$ represent
a regular expansion of the solution in power series in $\eps$ (an outer solution), and
$(u^{a}$, $v^{a}$, $p^{a})$ and $(u^{b}$, $v^{b}$, $p^{b})$  correspond to boundary layer corrections
to this regular expansion (inner solutions).
We assume that the boundary layer parts of the expansion rapidly decay outside thin boundary layers:
\begin{equation}
u^{a}, v^{a}, p^{a} =o(\xi^{-s}) \quad {\rm as} \quad \xi\to\infty \quad {\rm and} \quad
\quad u^{b}, v^{b}, p^{b} =o(\eta^{-s}) \quad {\rm as} \quad \eta\to\infty    \label{7}
\end{equation}
for every $s>0$. This assumption will be verified {\it a posteriori}.


\setcounter{equation}{0}
\renewcommand{\theequation}{3.\arabic{equation}}

\section{Asymptotic expansion}

\subsection{Regular part of the expansion}

Let
\begin{equation}
\bv^{r}=\bv^{r}_{0}+ \eps \, \bv^{r}_{1}+\eps^2 \bv^{r}_{2}+ \dots , \quad
p^{r}=p^{r}_{0}+ \eps \, p^{r}_{1}+\eps^2 p^{r}_{2}+ \dots , \label{3.1}
\end{equation}
where $\bv^{r}=(u^{r},v^{r})$, $\bv^{r}_{k}=(u^{r}_{k},v^{r}_{k})$ ($k=0,1,2,\dots$).
The successive approximations $\bv^{r}_{k}$, $p^{r}_{k}$ ($k=0,1,2,\dots$)
satisfy the equations:
\begin{equation}
\pr_{\tau}\bv^{r}_{k}=-\nabla p^{r}_{k}+\bF_{k}, \quad
\nabla\cdot\bv^{r}_{k}=0, \label{3.2}
\end{equation}
where $\bF_{0}\equiv 0$,  $\bF_{1}\equiv 0$ and
\begin{equation}
\bF_{k}=-\sum_{l=0}^{k-2}(\bv^{r}_{l}\cdot\nabla)\bv^{r}_{k-2-l}
+\nu\nabla^{2} \bv^{r}_{k-2} \label{3.3}
\end{equation}
for $k\geq 2$.
In what follows, we will use the following notation: for any $2\pi$-periodic $f(\tau)$,
\begin{equation}
f(\tau)=\bar{f}+\tilde{f}(\tau), \quad
\bar{f}=\frac{1}{2\pi}\int\limits_{0}^{2\pi}f(\tau)d\tau \label{3.4}
\end{equation}
where $\bar{f}$ is the mean value of $f(\tau)$ and, by definition, $\tilde{f}(\tau)=f(\tau)-\bar{f}$.

\subsubsection{Leading-order equations}

A general solution of Eqs. (\ref{3.2}) for $k=0$, which is periodic in $\tau$,
can be written as
\begin{equation}
\bv^{r}_{0}=\bar{\bv}^{r}_{0}+\tilde{\bv}_{0}, \quad \tilde{\bv}^{r}_{0}=\nabla\phi_{0} \label{3.5}
\end{equation}
where $\phi_{0}$ has zero mean value and is the solution of the boundary value problem
\begin{equation}
\nabla^{2}\phi_{0}=0, \quad
\phi_{0y}\vert_{y=0}=V^{a}_{0}(x,\tau), \quad  \phi_{0y}\vert_{y=1}=V^{b}_{0}(x,\tau),
\quad \phi_{0}(x+L,y)=\phi_{0}(x,y).  \label{3.6}
\end{equation}
The boundary conditions at $y=0$ and $y=1$ in (\ref{3.6}) will be justified later.

On averaging the equation for $\bv^{r}_{2}$ (the first equation (\ref{3.2}) for $k=2$) and the
incompressibility condition for $\bv^{r}_{0}$ and using the fact that $\tilde{\bv}^{r}_{0}$ is irrotational,
we obtain
\begin{equation}
(\bar{\bv}^{r}_{0}\cdot\nabla)\bar{\bv}^{r}_{0}
=-\nabla \Pi_{0}
+\nu\nabla^{2} \bar{\bv}^{r}_{0}, \quad
\nabla\cdot\bar{\bv}^{r}_{0}=0, \label{3.7}
\end{equation}
where $\Pi_{0}=\bar{p}^{r}_{2}+\overline{\vert\nabla\phi_{0}\vert^2/2}$.
Equations (\ref{3.7}) represent the time-independent Navier-Stokes equations.
It will be shown later that boundary conditions for $\bar{\bv}^{r}_{0}$ are
\begin{equation}
\bar{\bv}^{r}_{0}\vert_{y=0}=\bar{\bv}^{r}_{0}\vert_{y=1}={\bf 0}. \label{3.8}
\end{equation}
The only solution of (\ref{3.7}) that is periodic in $x$ and
satisfies (\ref{3.8}) is zero solution:
\begin{equation}
\bar{\bv}^{r}_{0}\equiv{\bf 0}. \label{3.9}
\end{equation}
This means that there is no steady streaming in the leading order of the expansion.

\subsubsection{First-order equations}

The solution of Eqs. (\ref{3.2}) for $k=1$ can be written as
\begin{equation}
\bv^{r}_{1}=\bar{\bv}^{r}_{1}+\tilde{\bv}_{1}, \quad \tilde{\bv}^{r}_{1}=\nabla\phi_{1} \label{3.10}
\end{equation}
where $\phi_{1}$ has zero mean value and is the solution of the boundary value problem
\begin{equation}
\nabla^{2}\phi_{1}=0, \quad
\phi_{1y}\vert_{y=0}=a_{1}(x,\tau), \quad  \phi_{1y}\vert_{y=1}=b_{1}(x,\tau), \quad
\phi_{1}(x+L,y)=\phi_{1}(x,y). \label{3.11}
\end{equation}
Functions $a_{1}(x,\tau)$ and $b_{1}(x,\tau)$ will be defined later.

To obtain equations for $\bar{\bv}^{r}_{1}(\bx)$, we
average the incompressibility condition for $\bar{\bv}^{r}_{1}$ and the equation for $\bv^{r}_{3}$
(the first equation (\ref{3.2}) for $k=3$) and then take account of Eq. (\ref{3.9}) and the fact that $\tilde{\bv}^{r}_{0}$ and
$\tilde{\bv}^{r}_{1}$ are both irrotational. This yields
\begin{equation}
{\bf 0}=-\nabla \Pi_{1}+\nu\nabla^{2} \bar{\bv}^{r}_{1}, \quad \nabla\cdot\bar{\bv}^{r}_{1}=0. \label{3.12}
\end{equation}
where $\Pi_{1}=\bar{p}^{r}_{3}+\overline{\left(\nabla\phi_{0}\cdot\nabla\phi_{1}\right)}$.
Thus, the first order averaged outer flow is described by the Stokes equations.
Boundary conditions for $\bar{\bv}^{r}_{1}$ will be specified later.

\subsection{Boundary layers near the walls at $y=0$ and $y=1$}

\subsubsection{Boundary layer equations}

To derive boundary layer equations
near the bottom wall, we ignore $u^{b}$, $v^{b}$ and $p^{b}$, because they
are supposed to be small everywhere except a thin boundary layer near $y=1$,
and assume that
\begin{equation}
u=u^{r}_{0}+u^{a}_{0}+ \eps (u^{r}_{1}+u^{a}_{1})+ \dots , \ \
v=v^{r}_{0}+ \eps (v^{r}_{1}+v^{a}_{0})+ \dots , \ \
p=p^{r}_{0}+p^{a}_{0}+ \eps (p^{r}_{1}+p^{a}_{1})+ \dots \quad \label{3.13}
\end{equation}
We substitute these
into Eq. (\ref{1})  and take into account that $u^{r}_{k}$, $v^{r}_{k}$, $p^{r}_{k}$
satisfy (\ref{3.2}). Then we make the change of variables
$y=\eps \, \xi$, expand every function of $\eps \, \xi$ in Taylor's series at $\eps=0$
and collect terms of the equal powers in $\eps$. As a result, we obtain
\begin{equation}
u^{a}_{k\tau}+p^{a}_{kx}-\nu u^{a}_{k\xi\xi}=F^{a}_{k}, \quad
p^{a}_{k\xi}=G^{a}_{k}, \quad u^{a}_{kx}+v^{a}_{k\xi}=0 \label{3.14}
\end{equation}
for $k=0,1,\dots$ In Eqs. (\ref{3.14}), functions
$F^{a}_{k}$ and $G^{a}_{k}$ depend on $\bv^{r}_{0}, \dots, \bv^{r}_{k-1}$,
$u^{a}_{0}, \dots, u^{a}_{k-1}$, $v^{a}_{0}, \dots, v^{a}_{k-1}$. For $k=0,1$, these are
given by
\begin{equation}
F^{a}_{0} = 0, \quad F^{a}_{1} = -V^{a}_{0}(x,\tau)u^{a}_{0\xi}, \quad
G^{a}_{0} = 0, \quad G^{a}_{1}=0. \label{3.15}
\end{equation}
A similar procedure leads to the equations of the boundary layer near the upper wall:
\begin{equation}
u^{b}_{k\tau}+p^{b}_{kx}-\nu u^{b}_{2\eta\eta}=F^{b}_{k}, \quad
p^{b}_{k\eta}=-G^{b}_{k}, \quad u^{b}_{kx}-v^{b}_{k\eta}=0 \label{3.16}
\end{equation}
for $k=0,1,\dots$. Functions $F^{b}_{k}$ and $G^{b}_{k}$ for $k=0,1$ are given by
\begin{equation}
F^{b}_{0} = 0, \quad F^{b}_{1} = V^{b}_{0}(x,\tau)u^{b}_{0\eta}, \quad
G^{b}_{0} = 0, \quad G^{b}_{1}=0. \label{3.17}
\end{equation}
Before we describe how the above boundary layer equations can be solved, we
need to discuss boundary conditions that must be satisfied by terms of each order in our asymptotic
expansion.

\subsubsection{Boundary conditions}

In view of (\ref{7}), we require that for every $s>0$ and for each $k=0,1,\dots$,
\begin{equation}
u^{a}_{k}, v^{a}_{k}, p^{a}_{k} =o(\xi^{-s}) \quad {\rm as} \quad \xi\to\infty \quad {\rm and} \quad
\quad u^{b}_{k}, v^{b}_{k}, p^{b}_{k} =o(\eta^{-s}) \quad {\rm as} \quad \eta\to\infty .    \label{3.18}
\end{equation}
Now we substitute (\ref{3.13}) into the first Eq. (\ref{4})
and collect terms of equal powers in $\eps$. This leads to the following boundary conditions
at $y=0$:
\begin{eqnarray}
&&u^{r}_{0}\!\Bigm\vert_{y=0}+ \, u^{a}_{0}\!\Bigm\vert_{\xi=0}=U^{a}_{0}(x,\tau), \quad
v^{r}_{0}\!\Bigm\vert_{y=0} = V^{a}_{0}(x,\tau), \label{3.19} \\
&&u^{r}_{k}\!\Bigm\vert_{y=0}+ \, u^{a}_{k}\!\Bigm\vert_{\xi=0}=U^{a}_{k}(x,\tau), \quad
v^{r}_{k}\!\Bigm\vert_{y=0}+ \, v^{a}_{k-1}\!\Bigm\vert_{\xi=0}=V^{a}_{k}(x,\tau) \label{3.20}
\end{eqnarray}
for $k\geq 1$.
Similarly, boundary conditions at $y=1$ are given by
\begin{eqnarray}
&&u^{r}_{0}\!\Bigm\vert_{y=1}+ \, u^{b}_{0}\!\Bigm\vert_{\eta=0}=U^{b}_{0}(x,\tau), \quad
v^{r}_{0}\!\Bigm\vert_{y=1} = V^{b}_{0}(x,\tau), \label{3.21} \\
&&u^{r}_{k}\!\Bigm\vert_{y=1}+ \, u^{b}_{k}\!\Bigm\vert_{\eta=0}=U^{b}_{k}(x,\tau), \quad
v^{r}_{k}\!\Bigm\vert_{y=1}+ \, v^{b}_{k-1}\!\Bigm\vert_{\eta=0}=V^{b}_{k}(x,\tau) \label{3.22}
\end{eqnarray}
for $k\geq 1$. Note that (\ref{3.19}) and (\ref{3.21}) justify boundary conditions for
$\phi_{0y}$ in (\ref{3.6}).


\subsubsection{Leading order equations}

{\it Boundary layer at $y=0$}. In the leading order, the boundary layer at $y=0$ is described
by (\ref{3.14}) for $k=0$.
The condition of decay at infinity (in variable $\xi$) for $p^{a}_{0}$ and the second Eq. (\ref{3.14}) imply that
$p^{a}_{0}\equiv 0$. Hence, the first Eq. (\ref{3.14}) reduces to the heat equation
\begin{equation}
u^{a}_{0\tau}=\nu u^{a}_{0\xi\xi}. \label{3.23}
\end{equation}
Boundary condition for $u^{a}_{0}$ at $\xi=0$ follows from (\ref{3.19}) and is given by
\begin{equation}
u^{a}_{0}\!\bigm\vert_{\xi=0}=U^{a}_{0}(x,\tau)-\phi_{0x}(x,0,\tau). \label{3.24}
\end{equation}
The periodic (in $\tau$) solution of Eq. (\ref{3.23}) must satisfy (\ref{3.24}) and the condition of decay
at infinity.

On averaging Eq. (\ref{3.23}), we find that $\bar{u}^{a}_{0\xi\xi}=0$.
The only solution of this equation that satisfies the decay condition at infinity and
the boundary condition $\bar{u}^{a}_{0}\vert_{\xi=0}=0$ (which follows from (\ref{3.24}) and
the fact that $\bar{U}^{a}_{0}=0$ and $\bar{\phi}_{0}=0$) is zero solution. Thus, in the leading order
the boundary layer  at $y=0$  is purely oscillatory: $\bar{u}^{a}_{0}\equiv 0$.

The normal velocity $v^{a}_{0}$ is determined from the third
Eq. (\ref{3.14}) for $k=0$:
\begin{equation}
v^{a}_{0}(x,\xi,\tau)=\pr_{x}\int\limits_{\xi}^{\infty}u^{a}_{0}(x,\xi',\tau)d\xi'. \label{3.25}
\end{equation}
Here the constant of integration is chosen so as to guarantee that $v^{a}_{0}\to 0$ as
$\xi\to\infty$. According (\ref{3.13}), $v^{a}_{0}(x,\xi,\tau)$ enters the first-order term in the expansion
and, therefore, $v^{a}_{0}\!\bigm\vert_{\xi=0}$ gives us the boundary condition for
the next approximation of the outer solution. Indeed, in view of (\ref{3.20}) (with $k=1$),
we have
$v^{r}_{1}(x,y,\tau)\!\bigm\vert_{y=0}=V^{a}_{1}(x,\tau)-v^{a}_{0}(x,0,\tau)$, so that
function $a_{1}(x,\tau)$ in problem (\ref{3.11}) is
$a_{1}(x,\tau)=V^{a}_{1}(x,\tau)-v^{a}_{0}(x,0,\tau)$.
Note also that Eq. (\ref{3.25}) implies that $\bar{v}^{a}_{0}\equiv 0$.


{\it Boundary layer at $y=1$}. Exactly the same arguments as above lead to the problem
\begin{eqnarray}
&&u^{b}_{0\tau}=\nu u^{b}_{0\eta\eta}, \label{3.26} \\
&&u^{b}_{0}\!\bigm\vert_{\eta=0}= U^{b}_{0}(x,\tau)-\phi_{0x}(x,1,\tau), \quad 
u^{b}_{0}\to 0 \ \ {\rm as} \ \ \eta\to\infty. \label{3.27}
\end{eqnarray}
Again, it follows from (\ref{3.26}) and (\ref{3.27}) that $\bar{u}^{b}_{0}\equiv 0$.
The normal velocity $v^{b}_{0}$ is given by
\begin{equation}
v^{b}_{0}(x,\eta,\tau)=-\pr_{x}\int\limits_{\eta}^{\infty}u^{b}_{0}(x,\eta',\tau)d\eta'. \label{3.28}
\end{equation}
Then it follows from (\ref{3.22}) (with $k=1$) that
$b_{1}$ in problem (\ref{3.11}) is
$b_{1}(x,\tau)=V^{a}_{1}(x,\tau)-v^{b}_{0}(x,0,\tau)$.
Again, (\ref{3.28}) implies that $\bar{v}^{b}_{0}\equiv 0$.


{\it Averaged outer flow}. On averaging
(\ref{3.19}) and (\ref{3.21}) and using the fact that both boundary
layers are purely oscillatory and our assumption
that $\bar{\bV}^{a,b}_{0}={\bf 0}$, we arrive at conclusion that
$\bar{\bv}^{r}_{0}\!\bigm\vert_{y=0}={\bf 0}$ and $\bar{\bv}^{r}_{0}\!\bigm\vert_{y=1}={\bf 0}$.
This justifies boundary conditions (\ref{3.8}) and our conclusion that {\it there is no steady streaming
in the leading order of the expansion}.


\subsubsection{First-order equations}

{\it Oscillatory outer flow}. 
Since functions $a_{1}(x,\tau)$ and $b_{1}(x,\tau)$, which appear
in problem (\ref{3.11}), are now known, the problem can be solved, thus giving us
$\tilde{\bv}^{r}_{1}(x,y,\tau)$.


{\it Boundary layer at $y=0$}.
Consider now Eqs. (\ref{3.14}) for $k=1$.
Again, the condition of decay at infinity for $p^{a}_{1}$ and the second Eq. (\ref{3.14})
imply that $p^{a}_{1}\equiv 0$. Hence, we have
\begin{equation}
u^{a}_{1\tau}=\nu u^{a}_{1\xi\xi}-V^{a}_{0}(x,\tau) \, u^{a}_{0\xi}. \label{3.29}
\end{equation}
Averaging this equation, we find that
$\nu \bar{u}^{a}_{1\xi\xi}=\overline{V^{a}_{0}(x,\tau)u^{a}_{0\xi}}$.
Integration over $\xi$ yields
\begin{equation}
\bar{u}^{a}_{1}=-\frac{1}{\nu}\int\limits_{\xi}^{\infty}
\overline{V^{a}_{0}(x,\tau)u^{a}_{0}(x,\xi',\tau)} \, d\xi'. \label{3.30}
\end{equation}
Here the constants of integration are chosen so as to satisfy the condition of decay at infinity.
According to (\ref{3.20}), we have
\begin{eqnarray}
&&\bar{u}^{r}_{1}\vert_{y=0}=-\bar{u}^{a}_{1}\vert_{\xi=0}=
\frac{1}{\nu}\int\limits_{0}^{\infty}\overline{V^{a}_{0}(x,\tau)u^{a}_{0}(x,\xi',\tau)} \, d\xi', \label{3.31} \\
&&\bar{v}^{r}_{1}\!\bigm\vert_{y=0}=0. \label{3.32}
\end{eqnarray}
The oscillatory part of ${u}^{a}_{1}$, as well as both averaged and oscillatory parts of ${v}^{a}_{1}$
can also be found but are not needed in what follows.


{\it Boundary layer at $y=1$}.
A similar analysis leads to
\begin{eqnarray}
&&\bar{u}^{b}_{1}=\frac{1}{\nu}\int\limits_{\eta}^{\infty}
\overline{V^{b}_{0}(x,\tau)u^{b}_{0}(x,\eta',\tau)} \, d\eta',  \label{3.33} \\
&&\bar{u}^{r}_{1}\vert_{y=1}=-
\frac{1}{\nu}\int\limits_{0}^{\infty}\overline{V^{b}_{0}(x,\tau)u^{b}_{0}(x,\eta',\tau)} \, d\eta', \label{3.34} \\
&&\bar{v}^{r}_{1}\!\bigm\vert_{y=1}=0. \label{3.35}
\end{eqnarray}


{\it Averaged outer flow}.
The first order averaged outer flow is described by the Stokes equations (\ref{3.12}) subject to
the boundary conditions
(\ref{3.31}), (\ref{3.32}), (\ref{3.34}) and (\ref{3.35}).
This boundary value problem describes a steady Stokes flow produced by
a given distribution of the velocity at the boundary. It will have a non-zero solution
provided that at least one of the integrals on the right sides of Eqs. (\ref{3.31}) and (\ref{3.34})
is non-zero. In this case, {\it steady streaming is the effect of first order in $\eps$}. This is in contrast
with steady streaming produced by vibrating impermeable walls where a non-zero averaged velocity appears in the second order of the expansion. Thus, the steady component of the flow in our problem is much stronger
than that in the case of  vibrating impermeable walls.

{\it Leading order asymptotic for the averaged flow.} To summarise, we have obtained
the first non-zero term in the expansion of the steady component of the flow. The
averaged velocity
field has the form
\[
\bar{u}=\eps\left(\bar{u}_{1}^{r} +\bar{u}_{1}^{a}+\bar{u}_{1}^{b}\right)+O(\eps^2), \quad
\bar{v}=\eps \, \bar{v}_{1}^{r} +O(\eps^2)
\]
where boundary layer contributions $\bar{u}_{1}^{a}$ and $\bar{u}_{1}^{b}$ are given by
Eqs. (\ref{3.30}) and (\ref{3.33}) and where $\bar{\bv}_{1}^{r}$ is the solution of the Stokes equations
that satisfies boundary conditions (\ref{3.31}), (\ref{3.32}), (\ref{3.34}) and (\ref{3.35}).
If we introduce the stream function for the averaged flow $\bar{\psi}$ defined by the standard relations
$\bar{u}=\bar{\psi}_{y}$ and $\bar{v}= - \bar{\psi}_{x}$, then
the corresponding expansion of $\bar{\psi}$ will have the form
\[
\bar{\psi}=\eps \, \bar{\psi}_{1}^{r} +O(\eps^2)
\]
where $\bar{\psi}_{1}^{r}$ is the stream function for $\bar{\bv}_{1}^{r}=(\bar{u}_{1}^{r}, \bar{v}_{1}^{r})$.
Note that boundary layer terms do not appear in the leading order of the expansion for $\bar{\psi}$.

We note also that, in our problem, there is no need to consider the Stokes drift as it is the effect
of higher order
in $\eps$.

\setcounter{equation}{0}
\renewcommand{\theequation}{4.\arabic{equation}}

\section{Examples}

Below we consider a few simple examples in which the velocity at the walls oscillates harmonically in time.
In all the examples below, the tangent velocity at the walls is zero ($U^{a}=U^{b}=0$),
and the normal velocity is non-zero ($V^{a}\neq 0$ and $V^{b}\neq 0$). Examples with zero
normal velocity and non-zero tangent velocity are not considered, as in this case, steady streaming appears in higher order
approximations (in the case of half space this problem 
had been treated by \cite{VV2008}).

\subsection{\it Example 1: Standing waves}

\begin{figure}
\begin{center}
\includegraphics*[height=6cm]{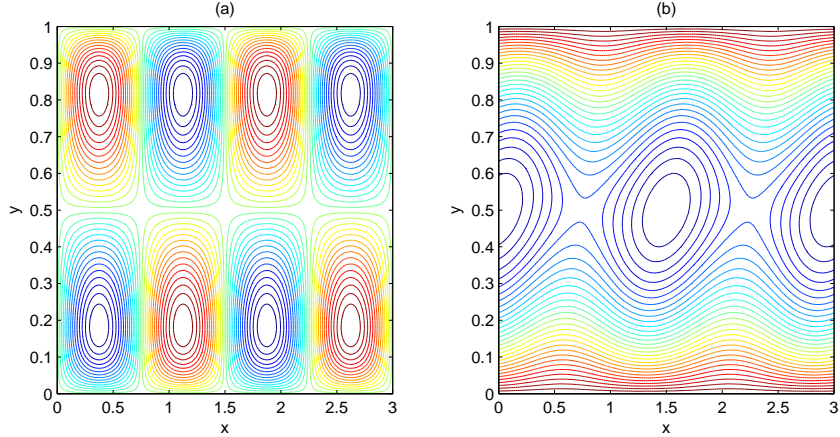}
\end{center}
\caption{The streamlines $\bar{\psi}_{1}^{r}=const$ for $\nu=1$, $L=3$ and $\alpha=1$:
(a) standing waves; (b) waves travelling in opposite directions.}
\label{fig1}
\end{figure}

Let $\bV^{a}=\cos k x \, \cos \tau \, \be_{y}$ and $\bV^{b}=\alpha \, \bV^{a}$
where $k=2\pi/L$ and $\alpha=\pm 1$. This choice corresponds to standing waves of injection/suction applied
at the boundaries of the channel ($\alpha=1$ if the waves are in phase, and $\alpha=-1$ if they have opposite phase).
After substitution of $\bV^{a}$ and $\bV^{b}$ in the general formulae of Section 3,
we find that
\begin{equation}
\bar{u}^{r}_{1}\vert_{y=0} = \bar{u}^{r}_{1}\vert_{y=1} =
-\frac{A(k)}{4\sqrt{2\nu}} \, \sin(2kx) , \quad 
A(k) \equiv \frac{\cosh(k)-\alpha}{\sinh(k)}. \label{4.3}
\end{equation}
Solving the Stokes equations (\ref{3.12}) with boundary conditions (\ref{3.32}), (\ref{3.35}) and (\ref{4.3}),
we obtain
\begin{equation}
\bar{\psi}_{1}^{r} =-
\frac{1}{4\sqrt{2\nu}} \, \frac{A(k)}{\sinh(2k)-2k} \,
\left\{(y-1) \, \sinh(2ky)+y \, \sinh\left[2k(1-y)\right]\right\} \, \sin(2kx) . \label{4.4}
\end{equation}
Typical streamlines are shown in Fig. \ref{fig1}(a). It is clear from (\ref{4.4}) that the flow picture is the same for $\alpha=1$ and $\alpha=-1$, the only difference is in the magnitude of the flow. The latter is determined by $A(k)$. For $\alpha=1$, $A$ is an increasing function,
$A(k)\sim k$ for $k\ll 1$ and $A(k) \to 1$ as $k\to\infty$. For $\alpha=-1$, $A$ is a decreasing function,
$A(k)\sim 1/k$ for $k\ll 1$ and $A(k) \to 1$ as $k\to\infty$. So, for all $k>0$, the magnitude of the steady streaming for $\alpha=-1$ is greater
than for $\alpha=1$. Thus, the steady streaming is stronger when standing waves of suction/injection applied at the walls have opposite phase.
Also, in this case the magnitude of the steady streaming increases with the wavelength of the waves.

\subsection{\it Example 2: Waves travelling in the same direction}

Let now the velocity at both walls be purely normal and have the form of waves travelling
in the direction of the $x$ axis: $\bV^{a}=\cos (k x-\tau) \, \be_{y}$ and $\bV^{b}=\alpha \, \bV^{a}$
where $k$ and $\alpha$ are the same as in Example 1.
Boundary conditions (\ref{3.31}) and (\ref{3.34}) take the form
\begin{equation}
\bar{u}^{r}_{1}\vert_{y=0} = \bar{u}^{r}_{1}\vert_{y=1} =
-\frac{A(k)}{2\sqrt{2\nu}} . \label{4.6}
\end{equation}
where $A(k)$ is given by (\ref{4.3}).
The Stokes equations (\ref{3.12}) subject to (\ref{3.32}), (\ref{3.35}) and (\ref{4.6})
lead to the constant solution\footnote{The Stokes equations (\ref{3.12}) also admit solutions
with a nonzero pressure gradient $\nabla \Pi_{1}=c_{0}\be_{x}$ ($c_{0}=const$), which are not considered here,
because this would be equivalent to a modification of our 
problem, allowing the presence of a weak $O(\eps^3)$ pressure gradient.}:
$\bar{u}^{r}_{1} = -A(k)/2\sqrt{2\nu}$, $\bar{v}^{r}_{1} = 0$. 
Thus, the waves travelling in the same direction produce a constant mean flow whose direction is opposite to the direction
in which the waves advance. This (somewhat surprising) result agrees with numerical simulations reported in \citep{Hoepffner}.
The magnitude of the mean flow is determined by $A(k)$. 
Properties of $A(k)$ imply that the most efficient way 
to generate the mean unidirectional flow is to apply suction/blowing in the form of waves
travelling in the same direction and having opposite
phase.

\subsection{\it Example 3: Waves travelling in the opposite directions}

Let now the normal velocity at the walls have the form of waves travelling in opposite
directions: $\bV^{a}=\cos (k x-\tau) \, \be_{y}$ and $\bV^{b}= \alpha \, \cos (k x+\tau) \, \be_{y}$
where $k$ and $\alpha$ are the same as in Examples 1 and 2.
Boundary conditions for $\bar{u}^{r}_{1}$ reduce to
\begin{equation}
\bar{u}^{r}_{1}\vert_{y=0} = -\frac{1}{2\sqrt{2\nu}} \, \frac{B^{-}(k,x)}{\sinh(k)}, \quad
\bar{u}^{r}_{1}\vert_{y=1} = \frac{1}{2\sqrt{2\nu}} \, \frac{B^{+}(k,x)}{\sinh(k)} \label{4.9}
\end{equation}
where $B^{\pm}=\cosh(k)+\alpha[\cos(2kx) \pm \sin(2kx)]$.
The corresponding solution of the Stokes equations is given by
\begin{equation}
\bar{\psi}^{r}_{1} = -\frac{1}{2\sqrt{2\nu}} \left[\frac{\cosh(k)}{\sinh(k)} \, y(1-y)+
 D^{-} \, y\sinh[2k(1-y)]+D^{+}(1-y)\sinh(2ky)\right] \label{4.10}
\end{equation}
where
\[
D^{\pm}(k,x)=\frac{\alpha}{\sinh(k)}\left[\frac{\cos(2kx)}{\sinh(2k)+2k} \pm \frac{\sin(2kx)}{\sinh(2k)-2k}\right].
\]
Typical streamlines of the flow (\ref{4.10}) for $\alpha=1$ are shown in Fig. \ref{fig1}(b). In the case of $\alpha=-1$,
the flow picture is the same except that it is shifted along $x$ axis by $L/4$ (this can be deduced directly from (\ref{4.10})).


\setcounter{equation}{0}
\renewcommand{\theequation}{5.\arabic{equation}}

\section{Conclusions}

We have considered incompressible flows in a channel between two parallel permeable walls
and constructed an asymptotic expansion of solutions of the Navier-Stokes equations
in the limit when the amplitude of displacements of fluid particles near the walls is mich smaller than both the width of the
channel and the thickness of the Stokes layer. The asymptotic procedure is based on the Vishik-Lyusternik method
and can be used to construct as many terms of the expansion as necessary.
In the leading order, the averaged flow is described by the stationary Stokes equations
subject to the boundary conditions that are determined by the boundary layers at the walls.
The key difference between the present expansion and the asymptotic theories of steady streaming induced
by vibrating impermeable boundaries is that, in our study, the magnitude of the steady streaming is $O(\eps)$,
which is much bigger than $O(\eps^2)$ steady streaming in the case of impermeable walls.

The general formulae have been applied to three particular examples of steady streaming induced by
blowing/suction at the walls in the form of standing and travelling plane waves.
In the case of standing waves,
the averaged flow has the form of a double array of vortices (see Fig. \ref{fig1}(a)). For short waves the vortices are concentrated near the walls,
while long waves produce vortices that fill the entire channel.

If the normal velocity at the walls have the form of plane harmonic waves which travel in the same direction,
the induced steady flow is a constant unidirectional
flow whose direction is opposite to the direction in which the waves travel.
This is different from the case of the steady streaming generated by vibrations of impermeable walls where
the induced flow had the same direction as the travelling wave. As far as we are aware, this was first observed
in numerical simulations performed by \citet{Hoepffner} .

If the normal velocities at the walls have the form of plane waves traveling in opposite directions,
the averaged flow is a superposition of a shear flow with a linear velocity profile and a periodic array of vortices (`cat's eyes')
in the center of the channel. When the wavelength is small, the vortices are weak. Their intensity and size monotonically grow with the wavelength
and eventually they fill the entire gap between the walls.

There are many open problems in this area. In particular, it is not clear how the present theory can be extended
to the case of $R_{s}\sim 1$. Although the problem does not involve a moving boundary which, in general, simplifies
things, there is a technical difficulty of a different sort. It is related to the leading order boundary layer equations for $R_{s}\sim 1$
which are difficult to solve analytically\footnote{Note that in the case of vibrating impermeable walls and $R_{s}\sim 1$,
the same equations (but with different boundary conditions) can be solved analytically \citep[see][]{IM2011}.}.
This is a subject of a continuing investigation.


\vskip 0.3cm
\noindent
{\bf Acknowledgments.} I am grateful to A. B. Morgulis and V. A. Vladimirov for stimulating discussions.




\end{document}